\documentclass[preprint,showpacs,preprintnumbers,eqsecnum,amsmath,amssymb]{revtex4}
\usepackage{epsf}
\usepackage{color}
\usepackage{dcolumn}
\usepackage{bm}
\everymath{\displaystyle}

\begin{document} 

\title{Dirac fermions in strong gravitational fields}

\author{\firstname{Yuri N.}~\surname{Obukhov}}
\email{obukhov@math.ucl.ac.uk} \affiliation{Department of
Mathematics and Institute of Origins, University College London,
Gower Street, London, WC1E 6BT, UK}

\author{\firstname{Alexander J.}~\surname{Silenko}}
\email{silenko@inp.bsu.by} \affiliation{Research Institute of
Nuclear Problems, Belarusian State University, Minsk 220080,
Belarus}

\author{\firstname{Oleg V.}~\surname{Teryaev}}
\email{teryaev@theor.jinr.ru} \affiliation{Bogoliubov Laboratory
of Theoretical Physics, Joint Institute for Nuclear Research,
Dubna 141980, Russia}

\date{file ``ostrong21.tex", \today}

\begin {abstract}
We discuss the dynamics of the Dirac fermions in the general strong 
gravitational and electromagnetic fields. We derive the general Hermitian
Dirac Hamiltonian and transform it to the Foldy-Wouthuysen representation 
for the spatially isotropic metric. The quantum operator equations of motion 
are obtained and the semiclassical limit is analyzed.
The comparison of the quantum mechanical and classical equations
shows their complete agreement. The helicity dynamics in strong
fields is discussed. Squaring the covariant Dirac equation
explicitly shows a similarity of the interactions of
electromagnetic and gravitational fields with a charged and
spinning particle.
\end{abstract}

 \maketitle

\section{Introduction}

The quantum and semiclassical theory of Dirac particles in the
gravitational field was studied in the numerous papers (see Refs.
\cite{Oliveira} to mention but a few). The main attention was usually paid 
to the dynamics of fermions in the weak gravitational fields when the 
components of the spacetime metric tensor do not significantly deviate from 
the flat Minkowski metric. The analysis of the strong gravitational fields 
is of interest for the investigation of the processes near black holes and 
massive compact astrophysical objects. They may be also applied to describe 
the spaces of variable geometry and dimension discussed recently
\cite{Fiziev:2010je,Mureika:2011bv,Sotiriou:2011xy}. 
Yet another application corresponds to the fermions (quarks) motion in 
rotating media which may be realized \cite{Rogachevsky:2010ys}
in heavy-ion collisions.  

In this paper, we apply the methods of the Foldy-Wouthuysen (FW)
transformation previously used for the analysis of the spin
dynamics in weak static and stationary gravitational fields
\cite{PRD,PRD2,Warszawa,OST} for the case of arbitrary strong
gravitational fields. We use the isotropic spatial coordinates and
systematically investigate, in particular, the spin dynamics of
Dirac fermions. Specifically, we derive the general result for the
angular velocity of spin precession and compare this quantum
mechanical result with classical predictions.

The paper is organized as follows. In Sec.~\ref{Hamiltonian} we
derive the general Hermitian Dirac Hamiltonian for a charged particle in
an arbitrary curved spacetime interacting with the electromagnetic fields.
In this general framework, we further specialize to the dynamics
of the spin and momentum of the particle in a spatially isotropic
metric. Sec.~\ref{FW1} presents the FW Hamiltonian and operator
equations of motion. The equations describing the dynamics of a
classical spin are obtained in Sec.~\ref{CS} and we compare them
with the corresponding quantum equations, testing the equivalence
principle. The validity of the latter as well as the similarity of
the spin interactions with the gravitational and with the
electromagnetic fields is also manifested in the analysis of
squared covariant Dirac equation performed in Sec.~\ref{squar}.
The evolution of the helicity in the strong fields is discussed in
Sec.~\ref{H}. The results obtained are summarized in Sec.~\ref{final}.

We denote world indices by Latin letters $i,j,k,\ldots = 0,1,2,3$
and reserve first Greek letters for tetrad indices,
$\alpha,\beta,\ldots = 0,1,2,3$. Spatial indices are denoted by
Latin letters from the beginning of the alphabet, $a,b,c,\ldots =
1,2,3$. The separate tetrad indices are distinguished by hats. As
usual, $\wedge$ and $^\ast$ denote the exterior product and the
Hodge operator, respectively.

\section{Spinning particle in curved spacetime}\label{Hamiltonian}

\subsection{General spacetime metric}

A classical spinning particle is characterized by its position in
spacetime, $x^i(\tau)$ which is a function of the proper time
$\tau$, and by the 4-vector of spin $S^\alpha$. The 4-velocity of
a particle $U^\alpha = e^\alpha_i dx^i/d\tau$ is normalized by the
condition $g_{\alpha\beta}U^\alpha U^\beta = c^2$ where
$g_{\alpha\beta} = {\rm diag}(c^2, -1, -1, -1)$ is the flat Minkowski metric. 
To describe spinning particles 
in flat and curved spacetimes (as well as in arbitrary curvilinear
coordinates), we use the tetrad $e^\alpha_i$. When the
gravitational field is absent, one can choose the Cartesian
coordinates and apply the coincidence of the holonomic orthonormal
frame with the natural one ($e^\alpha_i = \delta^\alpha_i$). For
an arbitrary spacetime metric, $g_{\alpha\beta}e^\alpha_ie^\beta_i
= g_{ij}$. Let $t$ be the time and $x^a~ (a=1,2,3)$ be spatial
local coordinates. The general form of the line element of an
arbitrary gravitational field can be given by
\begin{equation}\label{LT}
ds^2 = V^2c^2dt^2 - \delta_{\widehat{a}\widehat{b}}W^{\widehat
a}{}_c W^{\widehat b}{}_d\,(dx^c - K^ccdt)\,(dx^d - K^dcdt).
\end{equation}
The functions $V$ and $K^a$, as well as the components of the
$3\times 3$ matrix $W^{\widehat a}{}_b$ may depend arbitrarily on
$t,x^a$. The total number $1+3+9=13$ of these functions is larger
than the number of components of the spacetime metric. However, it
is obvious that the structure of (\ref{LT}) allows for a
redefinition $W^{\widehat a}{}_b\longrightarrow L^{\widehat a}
{}_{\widehat c}W^{\widehat c}{}_b$ with the help of an arbitrary
local rotation $L^{\widehat a}{}_{\widehat c}(t,x)\in SO(3)$.
Taking this freedom into account, we end up with exactly 10
independent variables that describe the general spacetime metric.

The off-diagonal metric components $g_{0a}$ are related to the
effects of rotation, as is well known. They arise when the
functions $K^a$ are nontrivial. For example, the {\it exact}
metric of the flat spacetime seen by an accelerating and rotating
observer is obtained for the special case \cite{HN}
\begin{equation}
V = 1 + {\frac {{\bm a}\cdot{\bm r}}{c^2}},\qquad W^{\widehat
a}{}_b = \delta^a_b,\qquad K^a =-{\frac 1c}\,(\bm\omega\times\bm
r)^a,\label{VWni}
\end{equation}
where ${\bm a}$ describes acceleration of the observer and
$\bm\omega$ is an angular velocity of a noninertial reference
system. Both are independent of the spatial coordinates, but may
depend arbitrarily on time $t$. The slowly rotating massive body
produces the gravitational field of similar form; far from the
source the Lense-Thirring \cite{LT} metric is described by
\begin{equation}
V = V(x),\qquad W^{\widehat a}{}_b = \delta^a_b\,W(x), \qquad K^a
={\frac 1c}\epsilon^{abc}\omega_b(x)x_c.\label{LTm}
\end{equation}
The functions $V(x^a), W(x^a), \bm{\omega}(x^a)$ can be recovered
from the Kerr metric in the isotropic coordinates in the limit of
$r\rightarrow\infty$:
\begin{eqnarray}\label{Vkerr}
V &=&\left(1 - {\frac {\mu}{2r}}\right) \left(1 + {\frac {\mu}{2r}}\right)^{-1}
- {\frac {\mu a^2 - 3\mu(\bm{a}\cdot\bm{n})^2}{2r^3}} + {\cal O}(a^2r^{-4}),\\
W &=& \left(1 + {\frac {\mu} {2r}}\right)^{2} + {\frac {\mu a^2 -
3\mu (\bm{a}\cdot\bm{n})^2}{2r^3}} + {\cal O}(a^2r^{-4}),\label{Wkerr}
\end{eqnarray}
and $K^a$ is given by the same equation (\ref{LTm}) with
\begin{equation}
\bm{\omega} = {\frac{2\mu c}{r^3}}\,\bm{a}\left(1 - {\frac
{3\mu}{r}} + {\frac {21\mu^2}{4r^2}}\right) + {\cal O}(a^3r^{-5}).\label{Okerr}
\end{equation}
Here $r := \sqrt{\bm{x}\cdot\bm{x}}$ and $\bm{n} = \bm{r}/r$. The
constant vector $\bm{a} = (0, 0, a)$ is the rotation parameter of
the Kerr solution and $\bm{a}\cdot\bm{n}=az/r$. Also, $\mu =
GM/c^2$; the total mass $M$ and the total angular momentum $\bm{J}
= Mc\bm{a}$ define the Kerr black hole uniquely. These equations
are obtained from the Arnowitt-Deser-Misner form \cite{ADM} of the
Kerr solution performed earlier by Hergt and Sch\"afer
\cite{hergt} after dropping the terms violating the isotropy.

In the weak field approximation, we have studied the dynamics of
quantum and classical spin for the cases (\ref{VWni}) and
(\ref{LTm}) in Ref.~\cite{OST}.

\subsection{Dirac fermions}

In order to discuss the Dirac spinors, we need the orthonormal
frames. The preferable choice \cite{OST} is the Schwinger gauge:
\begin{equation}\label{coframe}
e_i^{\,\widehat{0}} = V\,\delta^{\,0}_i,\qquad e_i^{\widehat{a}} =
W^{\widehat a}{}_b\left(\delta^b_i - cK^b\,\delta^{\,0}_i\right),\qquad a=1,2,3.
\end{equation}
Tetrad (\ref{coframe}) is characterized by the condition
$e_a^{\,\widehat{0}} =0, a=1,2,3$. The same is automatically true
for the inverse tetrad:
\begin{equation}\label{frame}
e_{\,\widehat{0}}^i = {\frac 1V}\left(\delta_{\,0}^i +
\delta_{\,a}^icK^a\right), \qquad e^i_{\widehat{a}} =
\delta_{\,b}^iW^b{}_{\widehat a},\qquad a=1,2,3,
\end{equation}
where the inverse $3\times 3$ matrix, $W^a{}_{\widehat
c}W^{\widehat c}{}_b = \delta_b^a$, is introduced.

The covariant Dirac equation for spin-1/2 particles in
gravitational and electromagnetic fields has the form
\begin{equation}
(i\hbar\gamma^\alpha D_\alpha - mc)\Psi=0,\qquad \alpha=0,1,2,3.
\label{Dirac0}
\end{equation}
The Dirac matrices $\gamma^\alpha$ are defined in local Lorentz
(tetrad) frames. The spinor covariant derivatives are given by
\begin{equation}
D_\alpha = e_\alpha^i D_i,\qquad D_i = \partial _i + {\frac {iq}{\hbar c}} 
\,A_i + {\frac i4}\sigma^{\alpha\beta}\Gamma_{i\,\alpha\beta}.\label{eqin2}
\end{equation}
Here $\Gamma_i{}^{\alpha\beta} = - \Gamma_i{}^{\beta\alpha}$ are
the Lorentz connection coefficients, $\sigma^{\alpha\beta} = \frac
i2\left(\gamma^\alpha \gamma^\beta
-\gamma^\beta\gamma^\alpha\right)$. The Dirac particle is
characterised by the electric charge $q$, and $A_i$ is the
4-potential of the electromagnetic field.

Eqs. (\ref{Dirac0}),(\ref{eqin2}) show that the gravitational and
inertial effects are encoded in coframe and connection (see the
relevant discussion in Refs. \cite{Ob1,Ob2} and references
therein). For the general metric (\ref{LT}) with the tetrad
(\ref{coframe}) we find explicitly
\begin{eqnarray}
\Gamma_{i\,\widehat{a}\widehat{0}} &=& {\frac
{c^2}V}\,W^b{}_{\widehat{a}} \,\partial_bV\,e_i{}^{\widehat{0}} -
{\frac cV}\,{\cal Q}_{(\widehat{a}\widehat{b})}
\,e_i{}^{\widehat{b}},\label{connection1}\\
\Gamma_{i\,\widehat{a}\widehat{b}} &=& {\frac cV}\,{\cal
Q}_{[\widehat{a} \widehat{b}]}\,e_i{}^{\widehat{0}} + \left({\cal
C}_{\widehat{a}\widehat{b} \widehat{c}} + {\cal
C}_{\widehat{a}\widehat{c}\widehat{b}} + {\cal
C}_{\widehat{c}\widehat{b}\widehat{a}}\right)
e_i{}^{\widehat{c}}.\label{connection2}
\end{eqnarray}
Here 
\begin{eqnarray}
{\cal Q}_{\widehat{a}\widehat{b}} &=&
g_{\widehat{a}\widehat{c}}W^d{}_{\widehat{b}} \left({\frac
1c}\dot{W}^{\widehat c}{}_d + K^e\partial_e{W}^{\widehat c}{}_d +
{W}^{\widehat c}{}_e\partial_dK^e\right),\label{Qab}\\
{\cal C}_{\widehat{a}\widehat{b}}{}^{\widehat{c}} &=&
W^d{}_{\widehat{a}}
W^e{}_{\widehat{b}}\,\partial_{[d}W^{\widehat{c}}{}_{e]}=- {\cal
C}_{\widehat{b}\widehat{a}}{}^{\widehat{c}},\qquad {\cal
C}_{\widehat{a} \widehat{b}\widehat{c}} =
g_{\widehat{c}\widehat{d}}\,{\cal C}_{\widehat{a}
\widehat{b}}{}^{\widehat{d}}.\label{Cabc}
\end{eqnarray}
The dot $\dot{\,}$ denotes the derivative with respect to the time
$t$. As one notices, ${\cal
C}_{\widehat{a}\widehat{b}}{}^{\widehat{c}}$ is nothing but the
anholonomity object for the spatial triad ${W}^{\widehat a}{}_b$.
The indices (that all run from 1 to 3) are raised and lowered with
the help of the spatial part of the flat Minkowski metric,
$g_{\widehat{a}\widehat{b}} = -\,\delta_{ab} = {\rm diag}(-1, -1, -1)$.

Dirac equation can be derived from the action
\begin{equation}
I = \int\,d^4x\,{\cal L},\qquad {\cal L} =
\sqrt{-g}\,L\label{action}
\end{equation}
with the Lagrangian
\begin{equation}\label{LD}
L = {\frac {i\hbar}{2}}\left(\overline{\Psi}\gamma^\alpha
D_\alpha\Psi - D_\alpha\overline{\Psi}\gamma^\alpha\Psi\right) -
mc\,\overline{\Psi}\Psi.
\end{equation}
As it is well known, the naive Hamiltonian form the Schr\"odinger
form of the Dirac equation is not Hermitian. The most
straightforward way to derive a Hermitian Hamiltonian is to
redefine the wave function. Substituting the tetrad and connection
into the Lagrangian density, we find explicitly
\begin{eqnarray}
I &=& \int\,dtd^3x\left[{\frac {i\hbar}{2}}\sqrt{-g}e_{\widehat{0}}^0\left(
\Psi^\dagger\partial_t\Psi - \partial_t\Psi^\dagger\Psi\right) - qA_0\sqrt{-g}
e_{\widehat{0}}^0\Psi^\dagger\Psi -mc^2\sqrt{-g}\Psi^\dagger\Psi\right.\nonumber\\
&& +\,{\frac {i\hbar}{2}}\sqrt{-g}e_{\widehat{0}}^a\left(\Psi^\dagger\partial_a
\Psi - \partial_a\Psi^\dagger\Psi\right) - {\frac qc}
\,\sqrt{-g}e_{\widehat{0}}^aA_a\Psi^\dagger\Psi\nonumber\\
&& +\,{\frac {i\hbar c}{2}}\sqrt{-g}e_{\widehat{a}}^b\left(\Psi^\dagger\alpha^a
\partial_b\Psi - \partial_b\Psi^\dagger\alpha^a\Psi\right) - q\,\sqrt{-g}
e_{\widehat{a}}^bA_b\Psi^\dagger\alpha^a\Psi\nonumber\\ \label{action1}
&& -\,\left. {\frac \hbar 4}\sqrt{-g}\epsilon_{\widehat{a}\widehat{b}\widehat{c}}
\Gamma_{\widehat{0}}{}^{\widehat{a}\widehat{b}}\Psi^\dagger\Sigma^a\Psi
+ {\frac {\hbar c} 4}\sqrt{-g}\epsilon_{\widehat{a}\widehat{b}}{}^{\widehat{c}}
\Gamma_{\widehat{c}}{}^{\widehat{a}\widehat{b}}\Psi^\dagger\gamma_5\Psi\right].
\end{eqnarray}
A direct check shows that the Schr\"odinger equation derived from
this action has a non-Hermitian Hamiltonian. This problem
disappears if we define a new wave function by
\begin{equation}
\psi = \left(\sqrt{-g}e_{\widehat{0}}^0\right)^{\frac
12}\,\Psi.\label{newpsi}
\end{equation}
Such a non-unitary transformation may be also interpreted in the
framework of the pseudo-Hermitian quantum mechanics
\cite{GorNezn,GorNeznnew} (cf. also \cite{lec}). Substituting
(\ref{newpsi}) into (\ref{action1}), and recalling (\ref{coframe})
and (\ref{frame}), we find the action
\begin{eqnarray}
I &=& \int\,dtd^3x\left[{\frac {i\hbar}{2}}\left(
\psi^\dagger\partial_t\psi - \partial_t\psi^\dagger\psi\right) -
qA_0\psi^\dagger\psi - mc^2V\psi^\dagger\psi\right.\nonumber\\
&& +\,K^a\left\{{\frac {i\hbar}{2}}\left(\psi^\dagger\partial_a\psi -\partial_a
\psi^\dagger\psi\right) - {\frac qc}\,A_a\psi^\dagger\psi\right\}\nonumber\\
&& +\,{\cal F}^b{}_a\left\{{\frac {i\hbar c}{2}}\left(\psi^\dagger\alpha^a
\partial_b\psi - \partial_b\psi^\dagger\alpha^a\psi\right)
- q\,A_b\psi^\dagger\alpha^a\psi\right\}\nonumber\\
&& \left. - {\frac {\hbar c} 4}\left(\Xi_a\psi^\dagger\Sigma^a\psi
+ \Upsilon\,\psi^\dagger\gamma_5\psi\right)\right].\label{action2}
\end{eqnarray}
Here $V = e_0^{\widehat{0}}$, ${\cal F}^b{}_a =
\sqrt{-g}e_{\widehat{a}}^b = VW^b{}_{\widehat a}$, and
\begin{equation}
\Upsilon = -V\epsilon^{\widehat{a}\widehat{b}\widehat{c}}\Gamma_{\widehat{a}
\widehat{b}\widehat{c}} = - V\epsilon^{\widehat{a}\widehat{b}\widehat{c}} 
{\cal C}_{\widehat{a}\widehat{b}\widehat{c}},\qquad \Xi_{\widehat{a}} =
{\frac Vc}\,\epsilon_{\widehat{a}\widehat{b}\widehat{c}}\,\Gamma_{\widehat{0}}
{}^{\widehat{b}\widehat{c}} = \epsilon_{\widehat{a}\widehat{b}\widehat{c}} 
\,{\cal Q}^{\widehat{b}\widehat{c}}.\label{AB}
\end{equation}
For the static and stationary rotating configurations, the
pseudoscalar invariant vanishes
($\epsilon^{\widehat{a}\widehat{b}\widehat{c}} {\cal
C}_{\widehat{a}\widehat{b}\widehat{c}} = 0$), and thus the
corresponding term was absent in the special cases considered
earlier \cite{OST}. But in general this term contributes to the
Dirac Hamiltonian.

Variation of the action (\ref{action2}) with respect to the
rescaled wave function yields the Dirac equation in Schr\"odinger
form $i\hbar\frac{\partial \psi} {\partial t}= {\cal H}\psi$. The
corresponding {\it Hermitian} Hamiltonian reads
\begin{eqnarray}
{\cal H} &=& \beta mc^2V + q\Phi + {\frac c 2}\left(\pi_b\,{\cal
F}^b{}_a \alpha^a + \alpha^a{\cal F}^b{}_a\pi_b\right)\nonumber\\
\label{Hamilton1} && +\,{\frac c2}\left(\bm{K}\cdot\bm{\pi} +
\bm{\pi}\cdot\bm{K}\right) + {\frac {\hbar c}4}\left(\Upsilon
\gamma_5 +\bm{\Xi}\cdot\bm{\Sigma}\right).
\end{eqnarray}
The kinetic momentum operator $\pi_i = i\hbar\partial_i- {\frac
qc}A_i= p_i- {\frac qc}A_i$ accounts of the interaction with the
electromagnetic field $A_i = (\Phi, A_a)$.

Hamiltonian (\ref{Hamilton1}) is one of our central results and
covers the general case of a spin-1/2 particle in an arbitrary
curved spacetime. Remarkably, the form of the Hamiltonian remains
the same even when the connection is non-Riemannian. If the
torsion $T^\alpha = d\vartheta^\alpha + \Gamma_\beta{}^\alpha\wedge
\vartheta^\beta $ is nontrivial, we just have to replace
\begin{equation}\label{torsion}
\Upsilon\rightarrow\Upsilon - \check{T}_{\widehat{0}},\qquad
{\Xi}_a\rightarrow{\Xi}_a + \check{T}_{\widehat{a}},
\end{equation}
where $\check{T}_\alpha\vartheta^\alpha =
{}^\ast(\vartheta_\alpha\wedge T^\alpha)$ is the axial covector of
torsion (i.e., the totally antisymmetric piece of torsion). It is
worthwhile to mention that the recent discussion \cite{GM} of the
Dirac fermions in an arbitrary gravitational field is very
different in that the non-Hermitian Hamiltonian is used in that
work, in deep contrast to the explicitly Hermitian one
(\ref{Hamilton1}).

Since the metric tensor is symmetric, it can be brought to the form diagonal 
\emph{in spatial coordinates}. For example, the Kerr metric in the both 
spherical and Boyer-Lindquist coordinates belongs to this form. 

A spatially diagonal metric tensor can often be reduced to an isotropic 
form by an appropriate transformation of \emph{spatial coordinates}. 
In spatially isotropic coordinates, $W^{\widehat a}{}_b=W\delta^{a}{}_b$ and 
the final form of the line element is defined by
\begin{equation}\label{Ltisotr}
ds^2 = V^2c^2dt^2 - W^2\,\delta_{ab}\,(dx^a - K^acdt)\,(dx^b - K^bcdt).
\end{equation}

Evidently, none of the two transformations changes the temporal
coordinate. Since only the spatial coordinates are transformed,
the transformations do not change the physical frame (particle
rest frame) used for a definition of the three-component physical spin.

For isotropic metric (\ref{Ltisotr}) with the Schwinger gauge, the
exact Hermitian Dirac Hamiltonian reads \cite{OST}
\begin{eqnarray}
{\cal H} &=& \beta mc^2V + {\frac c 2}\left[(\bm{\alpha}
\cdot\bm{p}){\cal F} + {\cal F}(\bm{\alpha}\cdot\bm{p})\right]\nonumber\\
&& +\,{\frac c2}\left(\bm{K}\cdot\bm{p} + \bm{p}\cdot\bm{K}\right)
+ {\frac {\hbar
c}4}\,(\bm{\nabla}\times\bm{K})\cdot\bm{\Sigma},\label{Hamiltonian1}
\end{eqnarray}
where
\begin{equation}\label{deff}{\cal F} = V/W.
\end{equation}
It covers the general case of a spin-1/2 particle in an isotropic
stationary metric.

\section{The Foldy-Wouthuysen representation}\label{FW1}

Let us now derive the FW Hamiltonian for stationary spatially
isotropic metric (\ref{Ltisotr}). We perform the FW transformation
of the exact Dirac Hamiltonian (\ref{Hamiltonian1}) with the help
of the method developed in Ref. \cite{PRA}. Since the
gravitational field is supposed to be strong, we do not make any
approximations for the functions $V, W, K^a$, and the only small
parameter is the Planck constant $\hbar$. In the FW Hamiltonian,
we retain all the terms of the zero and first orders in $\hbar$
and the leading terms of order of $\hbar^2$ nonvanishing in the
both nonrelativistic and weak field approximations.
These terms describe the gravitational contact (Darwin)
interaction \cite{PRD}. The computations are straightforward, and
the final FW Hamiltonian is given by
\begin{equation}
{\cal H}_{FW}={\cal H}_{FW}^{(1)}+{\cal H}_{FW}^{(2)},\label{eqFW}
\end{equation} 
where
\begin{widetext}
\begin{eqnarray}
{\cal H}_{FW}^{(1)}=\beta\epsilon' - \beta\frac{\hbar mc^4}{4}\left\{
\frac{1}{2{\epsilon'}^2+mc^2\{\epsilon',V\}},\left[\bm{\Sigma}\cdot(\bm\Phi
\times\bm{p})-\bm{\Sigma}\cdot(\bm{p}\times
\bm\Phi)+\hbar\bm\nabla\cdot\bm\Phi\right]\right\}\nonumber\\
+ \beta\frac{\hbar c^2}{16}\left\{ \frac{1}{\epsilon'
},\left[\bm{\Sigma}\cdot (\mbox{\boldmath ${\cal G}$}\times\bm p)-
\bm{\Sigma}\cdot(\bm{p}\times\mbox{\boldmath ${\cal G}$})
+\hbar\bm\nabla\cdot\mbox{\boldmath ${\cal G}$}\right]\right\},\label{eq7}
\end{eqnarray}
\begin{eqnarray}
{\cal H}_{FW}^{(2)}=\frac c2\left(\bm K\!\cdot\!\bm p+\bm p\cdot\!\bm K\right) 
+\frac{\hbar c}{4}\bm\Sigma\cdot(\bm\nabla\times\bm K)-\frac{\hbar
c^3}{16}\left\{\frac{1}{2{\epsilon'}^2+mc^2\{\epsilon',V\}},\left\{{\cal
F}^2,\bm{\Sigma}\cdot \bm Q\right\}\right\}, \label{eq7K}
\end{eqnarray}
\end{widetext}
\begin{equation}
\epsilon'=\sqrt{m^2c^4V^2+\frac12c^2\{{\cal F}^2,\bm p^2\}},\qquad
\mbox{\boldmath ${\cal G}$}=\bm\nabla({\cal F}^2),\qquad
\bm{\Phi}={\cal F}^2 \bm\nabla V,\label{eqa}
\end{equation}
\begin{equation}
\bm Q=\bm p\times\bm\nabla\left(\bm K\!\cdot\!\bm p+\bm
p\cdot\!\bm K\right) -\bm\nabla\left(\bm K\!\cdot\!\bm p+\bm
p\cdot\!\bm K\right)\times\bm p -\bm p\times(\bm
p\times(\bm\nabla\times\bm K))-((\bm\nabla\times\bm K)\times \bm
p)\times\bm p.\label{eqq}
\end{equation}
The equivalent forms of the latter quantity
\begin{eqnarray}
\bm Q&=&\bm p\times(\bm p\times(\bm\nabla\times\bm
K))+((\bm\nabla\times\bm K) \times\bm p)\times\bm p+2\bm
p\times(\bm p\cdot\bm\nabla)\bm K-2\delta^{ab}
(\nabla_a\bm K)p_b\times\bm p,\nonumber\\
\bm Q&=&\bm p\times\bm\nabla(\bm p\cdot\bm K)-\bm\nabla(\bm
K\cdot\bm p)\times \bm p +\bm p\times(\bm p\cdot\bm\nabla)\bm
K-\delta^{ab}(\nabla_a\bm K)p_b\times\bm p\label{eqeq}
\end{eqnarray}
may be also useful. Eqs. (\ref{eqFW})--(\ref{eqeq}) (which in the
limiting cases agree with our previous results obtained in Refs.
\cite{PRD,PRD2,OST}) belong to the principal results of this paper.

The dynamical equation for the spin is obtained from the
commutator of the FW Hamiltonian with the polarization operator
$\bm\Pi=\beta\bm\Sigma$ and is given by
\begin{equation}
\frac{d\bm \Pi}{dt}=\frac{i}{\hbar}[{\cal H}_{FW},\bm \Pi]=\bm\Omega^{(1)}
\times\bm \Sigma+\bm\Omega^{(2)}\times\bm\Pi,\label{spinmeq}
\end{equation}
where $\bm\Omega^{(1)}$ is the operator of angular velocity of
rotation of the spin in the static gravitational field,
\begin{equation}
\bm\Omega^{(1)}=-\frac{mc^4}{2}\left\{\frac{1}{2{\epsilon'}^2+mc^2\{\epsilon',
V\}},(\bm\Phi\times\bm{p}-\bm{p}\times\bm\Phi)\right\}\\
+ \frac{c^2}{8}\left\{ \frac{1}{\epsilon' },(\mbox{\boldmath${\cal G}$}
\times\bm p- \bm{p}\times\mbox{\boldmath ${\cal G}$})\right\},\label{eqol}
\end{equation}
and the contribution from the nondiagonal part of the metric is equal to
\begin{eqnarray}
\bm\Omega^{(2)} &=&\frac{c}{2} \bm\nabla\times\bm K - \frac{c^3}{8}\left\{
\frac{1}{2{\epsilon'}^2+mc^2\{\epsilon',V\}},\left\{{\cal F}^2,\bm
Q\right\} \right\}. \label{finalOmega}
\end{eqnarray}

The two different matrices appear on the right-hand side of Eq.
(\ref{spinmeq}) due to the fact that $\bm\Omega^{(1)}$ contains
the velocity operator rather than the momentum one. Since the
velocity operator is proportional to an additional $\beta$ factor
and is equal to $\bm v = \beta c\bm p/\epsilon$ for free
particles, the operator $\bm\Omega^{(1)}$, when expressed in terms
of $\bm v$, also acquires an additional $\beta$ factor \cite{OST}.

The corresponding semiclassical formulas describing the motion of
the average spin are then explicitly given by
\begin{eqnarray}
{\frac {d{\bm s}}{dt}} &=& \bm \Omega\times{\bm s}
= (\bm \Omega^{(1)}+\bm \Omega^{(2)})\times{\bm s},\label{ds}\\
\bm\Omega^{(1)} &=& \frac{mc^4}{\epsilon'(\epsilon'+mc^2V)}\,\bm{p}\times\bm\Phi
- \frac{c^2}{2\epsilon'}\,\bm{p}\times\mbox{\boldmath ${\cal G}$},\label{FO}\\
\bm\Omega^{(2)} &=&\frac{c}{2}\,\bm\nabla\times\bm K
-\frac{c}{4\epsilon'(\epsilon'+mc^2V)}\,{\cal F}^2\bm Q,\label{finalOmegase}
\end{eqnarray}
where, in the semiclassical limit,
\begin{equation}
\epsilon'=\sqrt{m^2c^4V^2+c^2\bm p^2{\cal F}^2},\qquad \bm Q =
2\bm p\times\bm\nabla(\bm p\cdot\bm K) +2\bm p\times(\bm
p\cdot\bm\nabla)\bm{K}.\label{eQ}
\end{equation}

FW Hamiltonian (\ref{eqFW}) can now be expressed in the simpler
form in terms of $\bm\Omega^{(1)},\bm\Omega^{(2)}$:
\begin{eqnarray}
{\cal H}_{FW} &=& \beta\epsilon'+\frac c2\left(\bm K\!\cdot\!\bm p
+ \bm p\cdot\!\bm K\right) +\frac\hbar2\bm\Pi\cdot\bm\Omega^{(1)}+
\frac\hbar2\bm\Sigma\cdot\bm\Omega^{(2)}\nonumber\\
&& -\,\beta\frac{\hbar^2mc^4}{4}\left\{\frac{1}{2{\epsilon'}^2 +
mc^2\{\epsilon',V\}},\bm\nabla\cdot\bm\Phi\right\} +
\beta\frac{\hbar^2 c^2}{16}\left\{ \frac{1}{\epsilon'
},\bm\nabla\cdot\mbox{\boldmath ${\cal G}$}\right\}.\label{Hamlt}
\end{eqnarray}

It is instructive to compare the classical and quantum
Hamiltonians of a spinning particle. In order to do this, one can
start from the classical Hamiltonian of a spinless particle
\cite{Cogn}:
\begin{equation}
{\cal H}_{class}=\left(\frac{m^2c^2 - \tilde{g}^{ab}\pi_a\pi_b}{{g}^{00}}
\right)^{1/2} + \frac{g^{0a}\pi_a}{{g}^{00}} + qA_0,\qquad \tilde{g}^{ab}
= g^{ab} - \frac{g^{0a}g^{0b}}{{g}^{00}}.\label{clCog}
\end{equation}
Substituting the components of the general metric (\ref{LT}), we
recast this into
\begin{equation}
{\cal H}_{class}=\sqrt{m^2c^4V^2 + c^2\delta^{cd}{\cal F}^a{}_c\,{\cal F}^b{}_d 
\,\pi_a\,\pi_b\,} + c{\bm K}\cdot{\bm\pi} + q\Phi.\label{clasp0}
\end{equation}
In order to take into account the spin correctly, this Hamiltonian
should be completed by the interaction term $\bm s\cdot\bm\Omega$,
along the lines of the general discussion of the Ref. \cite{PK}:
\begin{equation}
{\cal H}_{class}=\sqrt{m^2c^4V^2 + c^2\delta^{cd}{\cal
F}^a{}_c\,{\cal F}^b{}_d \,\pi_a\,\pi_b\,} + c{\bm K}\cdot{\bm
\pi} + q\Phi+ \bm s\cdot\bm\Omega.\label{clasnewadded}
\end{equation} 
In the general case, $\bm\Omega$ includes the both electromagnetic and 
gravitational contributions.

For the case of the stationary spatially isotropic metric (\ref{Ltisotr}), 
we obtain the resulting classical Hamiltonian (equal just to $cp_0$)
\begin{equation}
{\cal H}_{class}=\sqrt{m^2c^4V^2+c^2{\cal F}^2\bm p^2} + c{\bm
K}\cdot{\bm p} + \bm s\cdot\bm\Omega, \label{clasp}
\end{equation}
where terms dependent on the electromagnetic fields are omitted.
It is satisfactory to notice that the quantum Hamiltonians
(\ref{eqFW}), and (\ref{Hamlt}) agree completely with the
classical Hamiltonian (\ref{clasp}).

The similarity of the quantum and the classical Hamiltonians
naturally leads to the similarity of the quantum and classical
equations of motion which we are going to discuss now.

\section{Classical dynamics of spinning particles}\label{CS}

The dynamics of spinning particles in strong gravitational fields
is described by the generally covariant Mathisson-Papapetrou
\cite{Mathisson,Papapetrou} theory. A different approach was
developed by Khriplovich and Pomeransky \cite{PK} for the
noncovariant 3-dimensional spin defined in the particle rest
frame. In general, the analysis of the Mathisson-Papapetrou
equations is a difficult task and various approximation schemes
were developed for their solution. By neglecting the second order
spin effects, the Mathisson-Papapetrou system is reduced to
\cite{chicone}
\begin{eqnarray}\label{dP}
{\frac {DU^\alpha}{d\tau}} &=& f_{\rm m}^\alpha = -\,{\frac 1{2m}}
\,S^{\mu\nu}U^\beta R_{\mu\nu\beta}{}^\alpha,\\
{\frac {DS^\alpha}{d\tau}} &=& 0.\label{dS}
\end{eqnarray}
The Mathisson force $f_{\rm m}^\alpha$ in the right-hand side of
(\ref{dP}) depends on the curvature $R_{\mu\nu\beta}{}^\alpha$ of
spacetime. The physical spin $\bm{s}$ is defined in the rest frame
of a particle. Taking into account that the 4-velocity is
\begin{equation}
U^i = \gamma (e^i_{\widehat 0}+ v^a\,e^i_{\widehat a}),\label{UGadd}
\end{equation}
we can recast the Mathisson-Papapetrou system into the 3-vector form
\begin{eqnarray}
{\frac {d\bm{v}}{d\tau}} &=& \bm{\mathcal{E}} - {\frac {\bm{
\bm{v}(\bm{v}\cdot\bm{\mathcal{E}})}} {c^2}} +
\bm{v}\times\bm{\mathcal{B}} + {\frac 1\gamma}\bm{f}_{\rm m},\label{force}\\
{\frac {d{\bm s}}{d\tau}} &=& \bm\Omega\times\bm s,\qquad
\bm{\Omega} = -\bm{\mathcal{B}} + {\frac {\gamma}{\gamma +
1}}\,{\frac {\bm{v} \times\bm{\mathcal{E}}} {c^2}},\label{omgem}
\end{eqnarray}
by introducing the fields $\bm{\mathcal E}$ and $\bm{\mathcal B}$ \cite{PK},
\begin{equation}\label{gem}
- U^i\Gamma_{i\widehat{0}}{}^{\widehat{a}} = {\mathcal E}^a,\qquad
- U^i\Gamma_{i\widehat{b}}{}^{\widehat{a}} = \epsilon^a{}_{bc}{\mathcal B}^c.
\end{equation}

This represents a clear analogy between the gravitational and
electromagnetic fields that makes it possible to speak of the
gravitoelectromagnetic type effects. One should not confuse the
objects (\ref{gem}) with the usual gravitoelectromagnetic fields
\cite{gem1,gem2} that are defined in the weak-field approximation
and that satisfy the Maxwell-like dynamical equations which are
derived from Einstein's gravitational equation. In contrast, the
fields (\ref{gem}) arise in a purely kinematic context and are
defined for arbitrarily strong field configurations. Note also
that they depend on the velocity of the particle, unlike the usual
gravitoelectromagnetic fields \cite{gem1,gem2}. Nevertheless,
their application is quite helpful because the equations of motion
of particles and their spins, written in terms of $\bm{\mathcal
E}$ and $\bm{\mathcal B}$, look very similar to the corresponding
equations of motion of charged spinning particles in the
electrodynamics.

Let us calculate these fields explicitly. We start from Eqs. (\ref{connection1})
and (\ref{connection2}) and use the notation from Ref. \cite{OST}. 
Multiplying the connection coefficients (\ref{connection1})-(\ref{connection2})
by the 4-velocity (\ref{UGadd}), we find:
\begin{eqnarray}\label{UG1}
U^i\,\Gamma_{i{\widehat a}{\widehat b}} &=& {\frac {c\gamma}{V}}
\,{\cal Q}_{[{\widehat{a}}{\widehat{b}}]} + \gamma\left({\cal C}_{\widehat{a} 
\widehat{b}\widehat{c}} + {\cal C}_{\widehat{a}\widehat{c}\widehat{b}} +
{\cal C}_{\widehat{c}\widehat{b}\widehat{a}}\right)v^{\widehat{c}},\\
U^i\,\Gamma_{i{\widehat 0}{\widehat b}} &=& {\frac {c\gamma}V}
\,{\cal Q}_{(\widehat{b}\widehat{c})}v^{\widehat{c}} - {\frac {c^2\gamma}V}
\,W^c{}_{\widehat{b}}\,\partial_cV.\label{UG2}
\end{eqnarray}
As a result, for the general metric (\ref{LT}), the
gravitoelectric and gravitomagnetic fields (\ref{gem}) read
\begin{eqnarray}\label{ge}
{\mathcal E}_a &=& {\frac {\gamma}V}\left(c{\cal
Q}_{(\widehat{a}\widehat{b})}v^b -
c^2\,W^b{}_{\widehat{a}}\,\partial_bV\right),\\ \label{gm}
{\mathcal B}^a &=& {\frac {\gamma}V}\left(-\,{\frac c2}\,{\Xi}^a -
{\frac 12}\Upsilon\,v^a + \epsilon^{abc}V{\cal C}_{bc}{}^dv_d\right).
\end{eqnarray}

The physical spin precesses (\ref{omgem}) with the angular
velocity ${\bm\Omega}$ that can also be written explicitly in
terms of the connection \cite{PK,OST}
\begin{equation}
\Omega_{\widehat{a}} = \epsilon_{abc}\,U^i\left({\frac 12}\Gamma_i{}^{\widehat{c}
\widehat{b}} +{\frac {\gamma}{\gamma + 1}}\,\Gamma_{i\widehat{0}}{}^{\widehat{b}}
v^{\widehat{c}}/c^2\right).\label{OmegaG}
\end{equation}
As compared with the quantities $\bm\Omega^{(1)}$ and
$\bm\Omega^{(2)}$ which describe the precession of the quantum
spin using the coordinate time, $\Omega_{\widehat{a}}$ contains an
extra factor $dt/d\tau = U^0 =\gamma/ V$ in view of a different
parameterization using the proper time.

Let us calculate the classical precession angular velocity in the
Schwinger gauge explicitly. Substituting (\ref{UG1}) and
(\ref{UG2}) into (\ref{OmegaG}), we obtain the {\it exact
classical formula} for the angular velocity of the spin precession
in an arbitrary gravitational field:
\begin{eqnarray}
\Omega^{\widehat{a}} &=& {\frac {\gamma}{V}}\left({\frac 12}{\Upsilon} 
\,v^{\widehat{a}} - \epsilon^{abc}V{\cal C}_{{\widehat{b}}{\widehat{c}}}{}^d 
v_{\widehat{d}} + {\frac \gamma{\gamma + 1}}\epsilon^{abc}W^d{}_{\widehat{b}}
\,\partial_dVv_{\widehat{c}}\right.\nonumber\\
&& +\left.{\frac c2}\,{\Xi}^{\widehat{a}} - {\frac \gamma{\gamma +
1}} \epsilon^{abc}{\cal Q}_{(\widehat{b}\widehat{d})}{\frac
{v^{\widehat{d}} v_{\widehat{c}}} c}\right).\label{OmegaS}
\end{eqnarray}
The terms in the first line are linear in the 4-velocity of the
particle, whereas the terms in the second line contain the even
number of the velocity factors. For the special class of
stationary geometries (\ref{Ltisotr}), this general formula
reduces to the simpler expression:
\begin{eqnarray}
\bm{\Omega} &=& {\frac \gamma V}\left[- V\,\bm{v}\times
\bm{\nabla}\left(\frac1W\right) - {\frac \gamma {W(\gamma +
1)}}\,\bm{v}\times\bm{\nabla}\,V \right.\nonumber\\ &&\left.
+\,{\frac c2}\,\bm{\nabla}\times\bm{K} - {\frac \gamma {2c(\gamma
+ 1)}}\left(\bm{v}\times\bm{\nabla}(\bm{v}\cdot\bm{K})+\bm{v}\times
(\bm{v}\cdot\nabla)\bm{K}\right)\right]. \label{OmegaFv}
\end{eqnarray}
Equivalently,
\begin{eqnarray}
\bm{\Omega} &=& {\frac \gamma V}\left[{\frac 1 {{\cal F}V(\gamma +
1)}} \,\bm{v}\times\bm\Phi -{\frac {1} {2{\cal F}}}\,\bm{v}\times
\mbox{\boldmath ${\cal G}$}\right.\nonumber\\
&&\left. +\,{\frac c2}\,\bm{\nabla}\times\bm{K} -{\frac \gamma{2c(\gamma + 1)}}
\left(\bm{v}\times\bm{\nabla}(\bm{v}\cdot\bm{K})+\bm{v}\times
(\bm{v}\cdot\bm\nabla)\bm{K}\right)\right].\label{OmegaFe}
\end{eqnarray}

Since
\begin{eqnarray}
\epsilon' = mc^2V\sqrt{1 + \frac{\bm{p}^2}{W^2m^2c^2}} =
mc^2V\gamma,\qquad \bm{p} = mW\gamma\bm{v},\label{conne}
\end{eqnarray}
we conclude that the classical equation of the spin motion (\ref{omgem})
agrees with the quantum equation (\ref{spinmeq}) and with the
semiclassical one (\ref{ds}). Thus, the classical and the quantum
theories of the spin motion in gravity are in complete agreement.
This is now verified for the arbitrary strong field
configurations.

\section{Evolution of helicity in an isotropic metric}\label{H}

Let us analyse the semiclassical evolution of the helicity of a
particle propagating in a strong gravitation field. The helicity
describes the spin orientation with respect to the direction of
particle's motion. As usual, one means the orientation of the
3-component spin defined in the particle rest frame (physical
spin). The particle motion is defined by the trajectory that shows
how the contravariant spatial world coordinates of the particle
change in time. Evidently, the particle motion can be correctly
characterized by the evolution of the \emph{contravariant} world
velocity or the unit vector in its direction. As a result, one can
unambiguously define the helicity as a projection of the
3-component spin (pseudo)vector \emph{in the particle rest frame}
onto the direction of the unit vector along the contravariant
velocity \emph{in the world frame}. Thus, the helicity should be defined as
$$\zeta= (\bm s/s)\cdot(\bm U/U)=(\bm s/s)\cdot\bm{\mathcal{V}}/\mathcal{V},$$
where $\bm U=\{U^1,U^2,U^3\}$ and $\bm{\mathcal{V}}=\bm U/U^0$.
The investigation of the helicity is simplified when particle's
trajectory is infinite. In this case, one can apply the fact that
the vector $\bm N=\bm U/U$ coincides with the vectors ${\bm v}/v$
and ${\bm p}/p$ at the initial and final parts of such a
trajectory of the particle because of the very large distance to
the field source. Here $\bm
v=\{v^{\widehat{1}},v^{\widehat{2}},v^{\widehat{3}}\}$ is the
velocity in the anholonomic \emph{coframe} (\ref{UGadd}) and $\bm
p =\{-p_1,-p_2,-p_3\}$ is the \emph{covariant} momentum entering
the classical and quantum Hamiltonians. For the isotropic metric
under consideration,
$p_a=e_a^{\widehat{0}}p_{\widehat{0}}+e_a^{\widehat{b}}p_{\widehat{b}}=
Wp_{\widehat{a}}$ and
$p_a/p=Wv_{\widehat{a}}/(|W|v)=v_{\widehat{a}}/v$. As a result,
the change of the helicity on the whole trajectory can be given by
$\Delta\zeta'=\Delta\zeta$, where $\zeta'= (\bm s/s)\cdot\bm
n,~{\bm n} = {\bm v}/v={\bm p}/p$. Evidently, $\zeta'=\cos{\chi}$,
where $\chi$ is an angle between the $\bm s$ and $\bm n$ vectors.

In the general case, the evolution of the covariant 
momentum operator is defined by
\begin{eqnarray}
\frac{d\bm p}{dt}=\frac{i}{\hbar}[{\cal H}_{FW},\bm p].
\label{eqFi}\end{eqnarray} In the discussion of the semiclassical
dynamics of the particle, we can neglect the small spin-dependent force 
which enters into the equation of motion with an additional factor
$\hbar$. Thus, in the semiclassical approximation
$$\frac{dp^{-1}}{dt} = -\,\frac{1}{p^3}\,{\bm p}\cdot\frac{d{\bm p}}{dt},$$
and the dynamics of the unit vector $\bm n$ is given by
\begin{equation}
\frac{d\bm n}{dt} = -\,{\frac 1p}\,{\bm n}\times\left({\bm
n}\times \frac{d{\bm p}}{dt}\right).
\end{equation}
Therefore, the vector $\bm n$ rotates with the angular velocity
\begin{equation}\label{finalnf}
\tilde{\bm\omega} = {\frac 1p}\,{\bm n}\times \frac{d{\bm p}}{dt},
\end{equation}
and one can add a quantity $\kappa\bm n$ with an arbitrary factor
$\kappa$. Using the Eqs. (\ref{eqa}), (\ref{Hamlt}),
(\ref{conne}), (\ref{eqFi}), (\ref{finalnf}), we derive
\begin{equation}
\tilde{\bm\omega} = -\frac {c^2}{{\cal F}V\gamma^2v}\,{\bm n}\times\bm\Phi 
-\frac {v}{2{\cal F}}\,{\bm n}\times\mbox{\boldmath ${\cal G}$} 
-c\bm n\times\bm\nabla(\bm n\cdot\bm K).\label{finln}
\end{equation}
The nabla operator does not act on $\bm n$.

We can add the quantity $\frac c2\bm n(\bm n\cdot(\bm n\times\bm
K))$ to $\tilde{\bm\omega}$ and use the identity
$$2\bm n\times\bm\nabla(\bm n\cdot\bm K)=\bm n(\bm n\cdot(\bm n\times\bm K))
-\bm{\nabla}\times\bm{K}+\bm n\times\bm\nabla(\bm n\cdot\bm
K)+\bm{n}\times (\bm{n}\cdot\bm\nabla)\bm{K}.$$ As a result, Eq.
(\ref{finln}) is recast into
\begin{equation}
\tilde{\bm\omega} = -\frac {c^2}{{\cal F}V\gamma^2v}\,{\bm
n}\times\bm\Phi -\frac {v}{2{\cal F}}\,{\bm n}\times\mbox{\boldmath 
${\cal G}$}+\,{\frac c2}\,\bm{\nabla}\times\bm{K} - {\frac
c2}\left[\bm{n}\times\bm{\nabla}(\bm{n}\cdot
\bm{K})+\bm{n}\times(\bm{n}\cdot\bm\nabla)\bm{K}\right].\label{finlf}
\end{equation}

The angular velocity of the spin rotation defined by Eq. (\ref{finalOmegase}) 
can be expressed in the form similar to Eq. (\ref{OmegaFe}):
\begin{eqnarray}
\bm{\Omega} &=& {\frac v {{\cal F}V(\gamma + 1)}}
\,\bm{n}\times\bm\Phi -{\frac {v} {2{\cal F}}}\,\bm{n}\times
\mbox{\boldmath ${\cal G}$}\nonumber\\
&& +\,{\frac c2}\,\bm{\nabla}\times\bm{K} -{\frac {c(\gamma-1)}{2\gamma}}
\left[\bm{n}\times\bm{\nabla}(\bm{n}\cdot\bm{K})+\bm{n}\times
(\bm{n}\cdot\bm\nabla)\bm{K}\right]. \label{finalOmegred}
\end{eqnarray}
Therefore, the vector of spin rotates with respect to the momentum
direction, and the angular velocity of such rotation is
\begin{eqnarray}
\bm{o} =\bm{\Omega}-\tilde{\bm\omega} ={\frac {c^2} {{\cal
F}V\gamma v}} \,\bm{n}\times\bm\Phi +{\frac {c}{2\gamma}}
\left[\bm{n}\times\bm{\nabla}(\bm{n}\cdot\bm{K})+\bm{n}\times
(\bm{n}\cdot\bm\nabla)\bm{K}\right]. \label{finalOm}
\end{eqnarray}

The same result can be obtained for an arbitrary metric by the
purely classical method considered in the previous section.
Indeed, let us use classical equations (\ref{force}),(\ref{omgem})
and neglect the relatively small Mathisson force ${\bm f}_{\rm m}$. Since
\begin{equation}
\frac{d\bm n}{d\tau} = -\,{\frac 1v}\,{\bm n}\times\left({\bm
n}\times \frac{d{\bm v}}{d\tau}\right),
\end{equation}
Eq. (\ref{force}) leads to the following formula:
\begin{equation}
\frac{d\bm n}{d\tau} = {\bm\omega}'\times{\bm n},\label{finalff}
\end{equation}
where
\begin{equation}\label{finalno}
{\bm\omega}'=-\,\bm{\mathcal{B}}+{\frac 1 {v^2}}\,{\bm v}\times\bm{\mathcal{E}}.
\end{equation}
Consequently, we find that the spin vector rotates with respect to
the momentum direction and the angular velocity of this rotation is
\begin{equation}
{\bm o} = \frac{V}{\gamma}(\bm{\Omega}-\bm{\omega}') = -V\,{\frac {{\bm v}
\times\bm{\mathcal{E}}} {\gamma^2 v^2}}.\label{fino}
\end{equation}
The gravitomagnetic field $\bm{\mathcal{B}}$ does not influence
particle's helicity \cite{Teryaev:2003ch}. This conclusion is
valid for an arbitrary gravitational field when the effect of the
Mathisson force on the particle motion is neglected.
$\bm{\mathcal{E}}$ is proportional to $\gamma/V$. The use of the
world time shows that any gravitational field does not affect the
helicity of a particle with a negligible mass.

The covariant Dirac equation can be used and the FW transformation
can be performed for a massless particle. However, the
interpretation of the results is a serious problem. The
3-component spin is defined in the particle rest frame which
cannot be introduced for such a particle.

For stationary isotropic metric (\ref{Ltisotr}), the gravitoelectric and 
the gravitomagnetic fields (\ref{ge}) and (\ref{gm}) reduce to
\begin{eqnarray}\label{ige}
{\mathcal E}_a &=& -\,{\frac {\gamma c^2}V}\left({\frac
1W}\partial_aV + {\frac {K^b\partial_bW}{W}}\,{\frac {v_a}{c}} +
\partial_{(a}K_{b)} \,{\frac {v^b}{c}}\right),\\ \label{igm}
{\mathcal B}^a &=& {\frac {\gamma}V}\left(-\,{\frac c2}\,\epsilon^{abc}
\partial_bK_c + \epsilon^{abc}V\,{\frac {\partial_bW}{W^2}}\,v_c\right).
\end{eqnarray}
This demonstrates the perfect agreement between the quantum and
classical analyses, cf. Eqs. (\ref{finalOm}) and (\ref{fino}).

As a final remark, we should mention that the results presented in
this section cannot, generally speaking, be applied for the particle moving 
on a finite trajectory because the replacement of $\zeta$ by $\zeta'$ is
not always possible in this case.  

\subsection{General noninertial frame}

A general noninertial frame is an important application of the
results above. This frame is characterized by the acceleration
${\bm a}$ and the rotation ${\bm \omega}$ of an observer. The {\it exact} 
metric of the flat spacetime seen by the accelerated and rotating observer 
has the form (\ref{Ltisotr}) and is given by Eq. (\ref{VWni}).

We can apply our general results (\ref{spinmeq})--(\ref{finalOmega}), 
(\ref{Hamlt}) for the metric (\ref{VWni}) of the flat spacetime. 
For the FW Hamiltonian we then find
\begin{equation}
\begin{array}{c}
{\cal H}_{FW}={\cal H}_{0}+\frac\hbar2\bm\Pi\cdot\bm\Omega^{(1)}+
\frac\hbar2\bm\Sigma\cdot\bm\Omega^{(2)},\\
{\cal H}_{0}=\frac\beta2\left\{\left(1+\frac{\bm a\cdot\bm r}{c^2}\right),
\sqrt{m^2c^4+c^2\bm p^2}\right\}-\bm\omega\cdot\bm l,\\
\bm\Omega^{(1)}= \frac{\bm a\times\bm p}{mc^2(\gamma+1)},\qquad\bm\Omega^{(2)}
=-\bm\omega,\qquad\gamma=\frac{\sqrt{m^2c^4+c^2\bm p^2}}{mc^2}.
\end{array}\label{Hamltni}
\end{equation}
Here ${\bm l} = {\bm r}\times{\bm p}$ is the angular momentum
operator. There are no any terms of order of $\hbar^2$
nonvanishing in both the nonrelativistic and the weak field
approximations. We also emphasize the absence of the
spin-dependent Mathisson force. Let us stress that Eq.
(\ref{Hamltni}) is derived for the strong kinematical effects when
the ratios $|\bm a\cdot \bm r|/c^2$ and $|\bm\omega\times\bm r|/c$
are not small. Nevertheless, Hamiltonian (\ref{Hamltni}) coincides
in the case of $\bm\omega=0$ with the result \cite{PRD} obtained
for the special case of the accelerated frame in the weak field
approximation $|\bm a\cdot\bm r|/c^2\ll1$. When $\bm a=0$, we
reproduce the exact FW Hamiltonian deduced in Ref. \cite{PRD2} for
Dirac particles in the rotating frame. Eq. (\ref{Hamltni}) and the
corresponding Hamiltonian obtained in Ref. \cite{PRD2} generalize
the approximate expression derived by Hehl and Ni \cite{HN}.

The equivalent form of ${\cal H}_{0}$ reads
\begin{equation}
{\cal H}_{0} = \beta\left(mc^2 + m{\bm a}\cdot{\bm r}\right)\gamma
- {\frac {i\hbar\beta\,{\bm a}\cdot{\bm p}}{2mc^2\,\gamma}} -
{\bm\omega}\cdot{\bm l}.\label{H0a}
\end{equation}

The metric under consideration is non-Minkowskian at any parts of
finite and infinite trajectories of particles. Therefore, we need
to use the contravariant 4-velocity $U^i$ or the world velocity
$\bm{\mathcal{V}}=c\bm{U}/U^0$. Evidently,
$\bm{\mathcal{V}}/\mathcal{V}=\bm{U}/U$.
The world velocity operator is given by
\begin{equation}
\bm{\mathcal{V}}=\frac{d\bm r}{dt}=\frac{i}{\hbar}[{\cal
H}_{FW},\bm r]= \frac\beta2\left\{\left(1+\frac{\bm a\cdot\bm
r}{c^2}\right), \frac{c\bm p}{\sqrt{m^2c^2+\bm
p^2}}\right\}-\bm\omega\times\bm r. \label{eqwve}
\end{equation}
As a result, the semiclassical approximation for the world
acceleration operator
$$\bm w=\frac{d\bm{\mathcal{V}}}{dt}=\frac{i}{\hbar}[{\cal H}_{FW},
\bm{\mathcal{V}}]$$ reads
\begin{equation}
\bm{w}=-\bm a\left(1+\frac{\bm a\cdot\bm
r}{c^2}\right)-\bm\omega\times(2\bm{\mathcal{V}}
+\bm\omega\times\bm r)+\frac{(\bm a\cdot(2\bm{\mathcal{V}}
+\bm\omega\times\bm r))}{c^2+\bm a\cdot\bm
r}\left(\bm{\mathcal{V}} +\bm\omega\times\bm r\right).\label{eqwac}
\end{equation}
The exact semiclassical Eq. (\ref{eqwac}) agrees with the particular results 
obtained in Refs. \cite{PRD} and \cite{PRD2} (which are approximate for the
acceleration $\bm a$).

The expression for the angular velocity of rotation of the vector
$\bm N=\bm{\mathcal{V}}/\mathcal{V}$ is similar to Eq. (\ref{finalnf}):
\begin{equation}
\bm\omega_{\mathcal{V}} = {\frac {1}{{\mathcal{V}}}}\,{\bm
N}\times \frac{d{\bm{\mathcal{V}}}}{dt}.\label{finalmv}
\end{equation}
Explicitly,
\begin{equation}
\bm\omega_{\mathcal{V}} = -2\bm\omega+{\frac {\bm
N}{{\mathcal{V}}}}\times\left[ -\bm a\left(1+\frac{\bm a\cdot\bm
r}{c^2}\right)-\bm\omega\times(\bm\omega\times\bm r)+ \frac{(\bm
a\cdot(2\bm{\mathcal{V}} +\bm\omega\times\bm r))}{c^2+\bm
a\cdot\bm r}\left(\bm\omega\times\bm r\right)\right].\label{finaN}
\end{equation}

The exact semiclassical expression for the angular velocity of spin
rotation can be written in terms of $\bm{\mathcal{V}}$:
\begin{equation}
\bm\Omega=\frac{\bm a\times(\bm{\mathcal{V}}+\bm\omega\times\bm r)}{c^2 
\left[1+\frac{\bm a\cdot\bm r}{c^2}+\sqrt{\left(1+\frac{\bm a\cdot\bm
r}{c^2}\right)^2-\left(\frac{\bm{\mathcal{V}}+ \bm\omega\times\bm
r}{c}\right)^2}\right]}-\bm\omega.\label{HamlOme}
\end{equation}

To consider the helicity evolution, it is sufficient to examine
three specific cases. If $|\bm\omega\times\bm r|\ll c,~|\bm
a\cdot\bm r|\ll c^2$, one may retain only the terms linear in
$\bm\omega,\bm a$:
\begin{equation}
\bm o_{\mathcal{V}} =\bm\Omega-\bm\omega_{\mathcal{V}}= \bm\omega+
\frac{\bm{\mathcal{V}}\times\bm a}{\gamma{\mathcal{V}}^2}.\label{Hamlwfa}
\end{equation}

When $\bm a=0$ (a purely rotating frame), we find
\begin{equation}
\bm o_{\mathcal{V}} =\bm\omega+ \frac{\bm{\mathcal{V}}\times(\bm\omega\times
(\bm\omega\times\bm r))}{{\mathcal{V}}^2}.\label{Hamlwna}
\end{equation}
The presented solution for the rotating frame is exact.

When $\bm\omega=0$ (an uniformly accelerated frame), we obtain
\begin{equation}
\bm o_{\mathcal{V}} = \frac{\bm{\mathcal{V}}\times\bm
a}{\mathcal{V}^2} \left(1+\frac{\bm a\cdot\bm r} {c^2}\right)-
\frac{\bm{\mathcal{V}}\times\bm a}{c^2 \left[1+\frac{\bm a\cdot\bm
r} {c^2}+\sqrt{\left(1+\frac{\bm a\cdot\bm r}{c^2}\right)^2-
\frac{\mathcal{V}^2}{c^2}}\right]}.\label{Hamluaf}
\end{equation}

As follows from the Eq. (\ref{eqwve}), $\mathcal{V}/c=1+\bm
a\cdot\bm r/c^2$ for the ultrarelativistic particles with
negligible masses. For such particles, we find $\bm o_{\mathcal{V}} =0$.
Therefore, their helicity remains unchanged in the uniformly
accelerated frame.

We can conclude that the motion in the general noninertial
(arbitrarily rotating and accelerating) frame leads to the change
of the helicity even for the ultrarelativistic particle with a
negligible mass. A similar effect takes place in a gravitational
field when a particle trajectory is finite.

Another possible application of our general results is a rotating
massive thin shell which metric obtained by Brill and Cohen
\cite{BC} has the form (\ref{LTm}),(\ref{Ltisotr}). We will
analyse this elsewhere.

\section{Squared Dirac equation and the equivalence principle}\label{squar}

The electromagnetic and gravitational contributions to the
covariant derivative (\ref{eqin2}) manifest an obvious similarity
of the electromagnetic and gravitational effects. In this section
we further clarify this similarity by analysing the squared Dirac equation.

The commutator of the covariant derivative (\ref{eqin2}) reads
\begin{equation}
D_iD_j - D_jD_i = {\frac {iq}{\hbar c}}\,F_{ij} + {\frac
i4}\sigma_{\alpha\beta} R_{ij}{}^{\alpha\beta}.\label{DD}
\end{equation}
Here $F_{ij} = \partial_iA_j - \partial_jA_i$ is the electromagnetic field 
tensor, and $R_{ij}{}^{\alpha\beta}$ is the Riemann curvature tensor.

Acting with the conjugate Dirac operator $\left(i\hbar
\gamma^\alpha D_\alpha + mc\right)$ on (\ref{Dirac0}), we find the
squared Dirac equation
\begin{equation}
\left(- \hbar^2g^{ij}D_iD_j - \hbar^2\gamma^{[i}\gamma^{j]}D_iD_j
- m^2c^2\right) \psi = 0.\label{Dsq1}
\end{equation}
Substituting Eq. (\ref{DD}), we find
\begin{equation}
\left(- \hbar^2g^{ij}D_iD_j - {\frac {i\hbar q}{2c}}\,\gamma^{[i}\gamma^{j]}
F_{ij} + {\frac {\hbar^2}8}\gamma^{[i}\gamma^{j]}\gamma_{[\alpha}\gamma_{\beta]}
R_{ij}{}^{\alpha\beta} - m^2c^2\right)\psi = 0.\label{Dsq2}
\end{equation}
Expanding the product of gamma matrices, we derive the equation
\begin{equation}
\left(-\hbar^2g^{ij}D_iD_j - {\frac {\hbar q}{2c}}\,\sigma^{\alpha\beta}
F_{\alpha\beta} + {\frac {\hbar^2}4}\,R - m^2c^2\right)\psi = 0,\label{Dsq3}
\end{equation}
where $F_{\alpha\beta}=e_\alpha^i e_\beta^jF_{ij}$ are the
tensor-like electromagnetic field coefficients.

The special form of Eq. (\ref{Dsq3}) for the Dirac particle in a
gravitational field has been obtained in Ref. \cite{Christensenduff}.

Explicit computation yields
\begin{equation}\label{D2}
-\hbar^2g^{ij}D_iD_j = \pi^{i}\pi_{i} - \frac {\hbar}{4}\sigma^{\alpha\beta}
\left\{\pi^{i},\Gamma_{i\,\alpha\beta}\right\}+ \frac{\hbar^2}{16}
\sigma^{\alpha\beta}\sigma^{\mu\nu}\Gamma^i{}_{\alpha\beta}\Gamma_{i\,\mu\nu}.
\end{equation}

Since $$\sigma^{\alpha\beta}\sigma^{\mu\nu}\Gamma^i{}_{\alpha\beta}
\Gamma_{i\,\mu\nu}=\frac12\{\sigma^{\alpha\beta},\sigma^{\mu\nu}\}
\Gamma^i{}_{\alpha\beta}\Gamma_{i\,\mu\nu}=2\Gamma^i{}_{\alpha\beta}
\Gamma_i{}^{\alpha\beta}+i\varepsilon^{\alpha\beta\mu\nu}\Gamma^i{}_{\alpha\beta}
\Gamma_{i\,\mu\nu}\gamma_5,$$ we get finally
\begin{equation}
\left[\pi^{i}\pi_{i} - \frac {\hbar}{2}\sigma^{\alpha\beta}\left(\frac {q}{c} 
F_{\alpha\beta} + m\Phi_{\alpha\beta}\right) + {\frac {\hbar^2}4}\,R + {\frac
{\hbar^2}{16}}\left(2\Gamma^i{}_{\alpha\beta}\Gamma_i{}^{\alpha\beta}+ 
i\varepsilon^{\alpha\beta\mu\nu}\Gamma^i{}_{\alpha\beta}\Gamma_{i\,\mu\nu}
\gamma_5\right)- m^2c^2\right]\psi = 0,\label{Dfin}
\end{equation}
where 
\begin{equation} 
\Phi_{\alpha\beta}=\frac {1}{2m}\left\{\pi^{i}, \Gamma_{i\,\alpha\beta}\right\},
\qquad \gamma_5=-i\gamma^{\widehat{0}}\gamma^{\widehat{1}}\gamma^{\widehat{2}}
\gamma^{\widehat{3}}.\label{PKc}
\end{equation}

In the semiclassical approximation, $\pi^{i}=mU^{i}$, and
$\Phi_{\alpha\beta}$ coincides with the spin (and momentum)
transport matrix in a gravitational field (see Ref. \cite{OST})
and with the tensor-like coefficients $\gamma_{\alpha\beta
\lambda}u^{\lambda}$ \cite{PK}. It is analogous to the
electromagnetic field tensor and leads to the Dirac
gyro-gravitomagnetic ratio $g_{grav}=2$ in perfect agreement with
the equivalence principle which is also manifested in the
interaction of spin with gravity \cite{KO}. This means
\cite{Teryaev:2003ch} the absence of both the anomalous
gravitomagnetic moment and the gravitoelectric dipole moment which
are gravitational analogs of the anomalous magnetic moment and the
electric dipole moment, respectively.

Eq. (\ref{Dfin}) explicitly shows a similarity of the Dirac
particle interactions with electromagnetic and gravitational
fields. This similarity is caused by the similarity of the motion
of spinning particles in any external classical fields shown in
Ref. \cite{OST}.

\section{Conclusions}\label{final}

In this paper we have studied the quantum and classical dynamics
of spin $1/2$ fermions in the strong fields. We derived the
Hermitian Dirac Hamiltonian (\ref{Hamilton1}) that describes the
fermion in arbitrary electromagnetic and gravitational fields in
the Schwinger gauge. Applying the FW transformation, we
constructed the respective Hamiltonian (\ref{eqFW}) for an
arbitrary isotropic metric. This allowed us to obtain the operator
equation of spin precession (\ref{spinmeq}) and its semiclassical
limit (\ref{ds}). As a specific application, we proved that the
equations of motion in the uniformly accelerated rotating frame
obtained earlier for the small acceleration are in fact valid for
arbitrary large acceleration. We also derived the classical
equation (\ref{OmegaS}) for the spin precession in the general
gravitational field (in the Schwinger gauge) and observed its
consistency with the semiclassical limit of quantum equation, in
complete agreement with the equivalence principle. We confirmed
this conclusion by deriving and analyzing the squared Dirac
equation. The general evolution equations for the helicity were
obtained in quantum and classical pictures. Their application to
the dynamics in the general noninertial frame revealed the change
of the helicity even for the ultrarelativistic particle with a
negligible mass.

\section*{Acknowledgements}

We are indebted to V.P.~Neznamov, P.~Fiziev and D.V.~Shirkov for
stimulating discussions. This work was supported in part by JINR,
the Belarusian Republican Foundation for Fundamental Research
(Grant No. $\Phi$10D-001) and the Russian Foundation for Basic
Research (Grants No. 11-02-01538, 11-01-12103).

\end{document}